\documentclass[sigconf, authorversion=true, nonacm=true]{acmart}

\renewcommand\footnotetextcopyrightpermission[1]{} % removes footnote with conference information in first column
\def\BibTeX{{\rm B\kern-.05em{\sc i\kern-.025em b}\kern-.08emT\kern-.1667em\lower.7ex\hbox{E}\kern-.125emX}}
\usepackage{nicefrac}
\usepackage{siunitx}
\usepackage{array,framed}
\usepackage{booktabs}
\usepackage{
  color,
  soul,
  float,
  epsfig,
  wrapfig,
  graphicx
}
\usepackage{setspace}
\usepackage{latexsym,fancyhdr,url}
\usepackage{enumerate}
\usepackage{algorithm2e}
\usepackage{algpseudocode}
\usepackage{graphics}
\usepackage{xparse} % argument parsing -- \edist
\usepackage{xspace}
\usepackage{multirow}
\usepackage{csvsimple}
\usepackage{balance}
\usepackage{graphicx, caption, subcaption}

\usepackage{
  tikz,
  pgfplots,
  pgfplotstable
}
\usepackage{hyperref}

\usetikzlibrary{
  shapes.geometric,
  arrows,
  external,
  pgfplots.groupplots,
  matrix
}

\pgfplotsset{compat=1.9}
\usepackage{mathtools}

\usepackage{xcolor}
\DeclareMathAlphabet{\mathcal}{OMS}{cmsy}{m}{n}
\DeclareGraphicsExtensions{%
    .png,.PNG,%
    .pdf,.PDF,%
    .jpg,.mps,.jpeg,.jbig2,.jb2,.JPG,.JPEG,.JBIG2,.JB2}

\usepackage{xparse}
\newcommand{\bnm}{\begin{newmath}}
\newcommand{\enm}{\end{newmath}}

\newcommand{\bea}{\begin{eqnarray*}}%
\newcommand{\eea}{\end{eqnarray*}}%

\newcommand{\bne}{\begin{newequation}}
\newcommand{\ene}{\end{newequation}}

\newcommand{\bal}{\begin{newalign}}
\newcommand{\eal}{\end{newalign}}

\newenvironment{newalign}{\begin{align}%
\setlength{\abovedisplayskip}{4pt}%
\setlength{\belowdisplayskip}{4pt}%
\setlength{\abovedisplayshortskip}{6pt}%
\setlength{\belowdisplayshortskip}{6pt} }{\end{align}}

\newenvironment{newmath}{\begin{displaymath}%
\setlength{\abovedisplayskip}{4pt}%
\setlength{\belowdisplayskip}{4pt}%
\setlength{\abovedisplayshortskip}{6pt}%
\setlength{\belowdisplayshortskip}{6pt} }{\end{displaymath}}

\newenvironment{newequation}{\begin{equation}%
\setlength{\abovedisplayskip}{4pt}%
\setlength{\belowdisplayskip}{4pt}%
\setlength{\abovedisplayshortskip}{6pt}%
\setlength{\belowdisplayshortskip}{6pt} }{\end{equation}}

\newcounter{ctr}

\newcounter{mytable}
\def\mytable{\begin{centering}\refstepcounter{mytable}}
\def\endmytable{\end{centering}}

\newcounter{myfig}
\def\myfig{\begin{centering}\refstepcounter{myfig}}
\def\endmyfig{\end{centering}}

\newlength{\saveparindent}
\newlength{\saveparskip}
\newcommand{\E}{{\rm I\kern-.3em E}}

\newcommand{\figref}[1]{\mbox{Figure~\ref{#1}}}

\renewcommand{\eqref}[1]{\mbox{Equation~(\ref{#1})}}

\newcommand{\tabref}[1]{\mbox{Table~\ref{#1}}}

\def \part {part}

\renewcommand{\paragraph}[1]{\vspace*{6pt}\noindent\textbf{#1}\;}

\def \blackslug{\hbox{\hskip 1pt \vrule width 4pt height 8pt
    depth 1.5pt \hskip 1pt}}
\def \qed{\quad\blackslug\lower 8.5pt\null\par}
\newcounter{mynote}[section]

\newcommand\ignore[1]{}

\newcounter{rcnote}[section]

\newcounter{mrnote}[section]

\newcounter{fknote}[section]

\newcounter{anote}[section]

\DeclareMathSymbol{\mlq}{\mathord}{operators}{``}
\DeclareMathSymbol{\mrq}{\mathord}{operators}{`'}

\newcommand{\rhf}[2]{R_{f, \gamma}}

\DeclareDocumentCommand{\edist}{o o}{
  \ensuremath{
    \IfNoValueTF{#1}{{d}}{{\sf d}(#1,#2)}
  }
}

\newcommand{\olrk}[1]{\ifx\nursymbol#1\else\!\!\mskip4.5mu plus 0.5mu\left(\mskip0.5mu plus0.5mu #1\mskip1.5mu plus0.5mu \right)\fi}

\NewDocumentCommand{\indseq}{ O{1} O{r} }{{#1}\ldots {#2}}

\begin{document}
%\fontfamily{lmr}\selectfont
% \def\thetitle{A Practical Way to Generate Strong Keys from Noisy Data}
\title{Exploration of Time Reversal for\\Wireless Communications within Computing Packages}
%\fancyhead{}
%\def\thetitle{Towards Spatial Multiplexing in\\Wireless Networks within Computing Packages}
%\title{\thetitle}

\author{Ama Bandara}
\authornote{Both authors contributed equally to this research.}
\affiliation{
\institution{NaNoNetworking Center in Catalunya (N3Cat)\\Universitat Polit\`{e}cnica de Catalunya}
  %\streetaddress{8600 Datapoint Drive}
  \city{Barcelona}
  %\state{Texas}
  \country{Spain}
  %\postcode{78229}
  }
\email{ama.peramuna@upc.edu}

\author{F\'atima Rodr\'iguez-Gal\'an*}
\affiliation{
\institution{NaNoNetworking Center in Catalunya (N3Cat)\\Universitat Polit\`{e}cnica de Catalunya}
  %\streetaddress{8600 Datapoint Drive}
  \city{Barcelona}
  %\state{Texas}
  \country{Spain}
  %\postcode{78229}
  }
\email{fatima.yolanda.rodriguez@upc.edu}

\author{Elana Pereira de Santana}
\affiliation{
 \institution{Institute of High Frequency and Quantum Electronics\\University of Siegen}
 \city{Siegen}
% \state{Arunachal Prades}
 \country{Germany}}
\email{Elana.PSantana@uni-siegen.de}

\author{Peter Haring Bol\'ivar}
\affiliation{
 \institution{Institute of High Frequency and Quantum Electronics\\University of Siegen}
 \city{Siegen}
% \state{Arunachal Prades}
 \country{Germany}}
\email{peter.haring@uni-siegen.de}

\author{Eduard Alarc\'on}
\affiliation{
 \institution{NaNoNetworking Center in Catalunya (N3Cat)\\Universitat Polit\`{e}cnica de Catalunya}
  %\streetaddress{8600 Datapoint Drive}
  \city{Barcelona}
  %\state{Texas}
  \country{Spain}
  %\postcode{78229}
  }
  \email{eduard.alarcon@upc.edu}
 
\author{Sergi Abadal}
\affiliation{
 \institution{NaNoNetworking Center in Catalunya (N3Cat)\\Universitat Polit\`{e}cnica de Catalunya}
  %\streetaddress{8600 Datapoint Drive}
  \city{Barcelona}
  %\state{Texas}
  \country{Spain}
  %\postcode{78229}
  }
  \email{abadal@ac.upc.edu}

\begin{comment}
\author{F\'atima Rodr\'iguez-Gal\'an*, Elana Pereira de Santana$^\dagger$, Peter Haring Bol\'ivar$^\dagger$, Sergi Abadal*, \\Eduard Alarc\'on*}
\affiliation{*NaNoNetworking Center in Catalunya (N3Cat)\\Universitat Polit\`{e}cnica de Catalunya, Barcelona, Spain\\$^\dagger$Institute of High Frequency and Quantum Electronics\\University of Siegen, Siegen, Germany
}
\end{comment}

\date{}

\renewcommand{\shortauthors}{Bandara \textit{et al.}}
\renewcommand{\shorttitle}{Time Reversal within Computing Packages}
\begin{abstract}
Wireless Network-on-Chip (WNoC) is a promising paradigm to overcome the versatility and scalability issues of conventional on-chip networks for current processor chips. However, the chip environment suffers from delay spread which leads to intense Inter-Symbol Interference (ISI). This degrades the signal when transmitting and makes it difficult to achieve the desired Bit Error Rate (BER) in this constraint-driven scenario. Time reversal (TR) is a technique that uses the multipath richness of the channel to overcome the undesired effects of the delay spread. As the flip-chip channel is static and can be characterized beforehand, in this paper we propose to apply TR to the wireless in-package channel. We evaluate the effects of this technique in time and space from an electromagnetic point of view. Furthermore, we study the effectiveness of TR in modulated data communications in terms of BER as a function of transmission rate and power. Our results show not only the spatiotemporal focusing effect of TR in a chip that could lead to multiple spatial channels, but also that transmissions using TR outperform, BER-wise, non-TR transmissions it by an order of magnitude.

\end{abstract}

\ccsdesc[500]{Hardware~Radio frequency and wireless interconnect}
\keywords{Wireless-Network-on-Chip; Flip-chip; Time Reversal}

\maketitle

% Section I
\section{Introduction}
\label{sec:intro}
%%%%%%%%%%%%%%%%%%%%%
Modern technological advancements and the end of Moore's Law have driven the field of computer architecture towards heterogeneous Systems-in-Package (SiP), systems hosting an increasing number of CPU, GPU, and specialized accelerator chips \cite{naffziger2021pioneering, Radway2021}. These chips are interconnected through an interposer or a Printed Circuit Board (PCB) and, given the nature of such SiPs, such interconnect needs to support high bandwidth and low latency through multiple parallel signal transmissions within and across the chips.

NoCs and NiP arise to improve the interconnections demands in computing systems \cite{bertozzi2015fast}. However, conventional wired NoC/NiP approaches suffer from latency and efficiency issues when scaling to a large number of cores/chips and are unable to provide flexibility to manage variable traffic loads or heterogeneous computing technologies \cite{bertozzi2015fast, ganguly2022interconnects}. This paved the way for the WNoC paradigm as a potential complement to existing NoCs. 

WNoC is a promising technology that takes advantage of electromagnetic (EM) nano-communication \cite{Kodi2015, Shamim2017}. It proposes to have wireless links connecting distant processors across chips and making use of the computing package as propagation medium. The information reaches the destination in a single hop and with latencies of a few nanoseconds irrespective of the communication distance, as opposed to the multi-hop nature of the wired NoCs/NiPs. This allows to relieve or even eliminate the communication bottleneck imposed by the prevailing NoC/NiP communication architectures, promising to accelerate future computing systems \cite{guirado2021dataflow}. %All of this allows us to avoid the communication bottleneck within the chip environment with low latency in comparison with the prevailing NoC communication architecture.

This diminished latency delivered by WNoC comes with a price since wireless bandwidth is limited and needs to be shared among the cores. As a result, Medium Access Control (MAC) protocols and multiplexing schemes are required to avoid collisions and interference \cite{abadal2022graphene, rodriguez2022towards}. Also, at the physical layer, WNoC needs to adapt to chip resource constraints. The use of mm-wave and terahertz bands allows the integration of tens of antennas within each chip, whereas simple modulations such as On-Off Keying (OOK) are adopted to avoid bulky or power-hungry components at the transceiver \cite{Yi2021}. However, with such low order modulations, high symbol rates are needed to reach the 10+ Gb/s speeds expected for WNoC. 

The energy propagation around the computing package, which can act as a sort of reverberation chamber, leads to a high delay spread (DS) \cite{Rayess2017, chen2019channel, imani2021smart}. This quickly leads to inter-symbol interference (ISI) as we gradually increase the symbol rate to achieve the aforementioned 10+ Gb/s speeds. Therefore, the stringent Bit Error Rate (BER) requirements of this scenario ($10^{-15}$, similar to the error rate of a wire) are proving difficult to meet \cite{timoneda2020engineer}. Moreover, in considering multi-node communication with simultaneous signal transmissions as a solution for bandwidth sharing, each node will be superimposed with another EM wave which could evidently hinder the demodulation of the intended signal and disrupt the communication. It is difficult to claim traditional wireless interference mitigation techniques \cite{OFDM,ZFBF} introduced up to 5G wireless communication protocol stack (unless with modified hybrid architectures) in considering the more realistic signal transmission within the chip to avoid complex transceivers with high power consumption. Hence, a fundamental challenge in WNoC resides in the mitigation of interference and DS effects in reverberating channels towards the creation of multiple parallel high-speed links.

\begin{figure}[!t]
\centering
%%\vspace{-0.2cm}
\includegraphics[width=1\columnwidth]{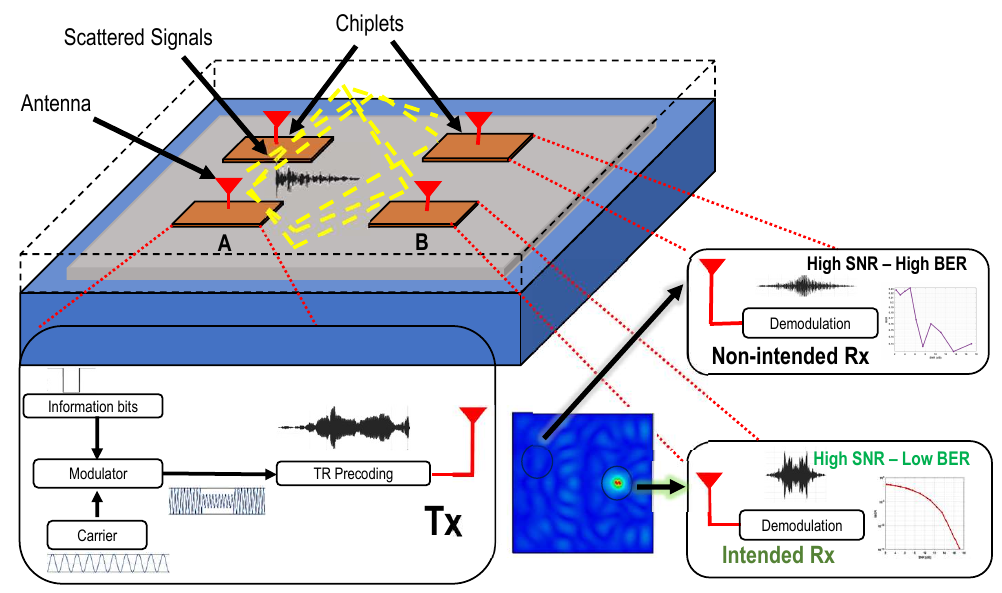}
\vspace{-0.4cm}
\caption{Time reversal in a computing package. Through a filter at the transmitter end, time reversal attains spatiotemporal focusing at an intended receiver, increasing the achievable data rate and opening the door to multiple spatial channels for wireless communications within a computing package.}
\label{fig:abstract}
\vspace{0.4cm}
\end{figure}

To bridge this gap, this paper proposes the use of Time Reversal in the wireless in-package environment (Figure \ref{fig:abstract}). TR is a technique that employs knowledge of the wireless channel impulse response to create a ideal matched filter \cite{lerosey2005time}. The key of TR is that it can be applied as a filter at the transmitter and, by doing so, the radiated signal is focused in time (with a peak much shorter than the length of the impulse response) and in space (around a targeted receiver). Hence, TR only mitigates potential ISI effects, but also compresses undesirable interference on adjacent transceivers paving the way to having multiple parallel spatial channels. 

%\hl{One paragraph summarizing very briefly TR in other environments (just cite), their problems and how we are better/opportunistic.}
The benefit of TR in rich scattering environments is well understood and proven. Originally demonstrated in acoustics, TR was then experimentally validated for wireless communications at microwave frequencies \cite{lerosey2004time} and, more recently, sub-terahertz bands \cite{mokh2022experimental, highfreqTR}. Various modulation schemes have been combined with TR with the aim of transmitting faster than what the DS of the channel would advise \cite{TRPPM1,TRPPM2}. However, since conventional wireless networks typically deal with dynamic channel variations, it is difficult to adopt TR due to the need of constantly obtaining the Channel State Information (CSI) and implementing an filter that can adapt to changes in the CSI.

%In current literature \cite{lerosey2004time}, it is proven with various experimental setups, better spatial and temporal convergence can be achieved by TR in rich scattering environments. Few experiments have also been conducted on sub terahertz frequency band \cite{mokh2022experimental, highfreqTR} to explore the effect of TR with high frequency communications in closed chambers. Various modulation schemes \cite{TRPPM1,TRPPM2} have been accessed with TR to evaluate the performance with the aim of transmitting more symbols with focused energy leading to higher BER.

%\hl{One paragraph detailing the contribution and main results achieved.}
In this paper, we present a simplified version of TR for WNoC. With the leverage of having prior knowledge of wireless channels within package, which are static, the TR filter(s) can be pre-designed and applied without the need for frequent CSI calculation. To show the proposed approach, we obtain the channel impulse response of a wireless in-package link in CST Microwave Studio and evaluate the spatiotemporal focusing ability of TR in this scenario. We then apply different modulation schemes and analyze the BER as a function of data rate and Signal-to-Noise Ratio (SNR). Compared to a non-TR approach, we show order-of-magnitude improvements for both interference suppression and achievable data rate.

%The conventional wireless transmission is constantly subjected to dynamic channel variation with small scale fading and it is difficult to adopt TR technique to gain the best outcome with random CSI obtained in short time intervals. With the leverage of having static and quasi deterministic wireless channels in flip-chip environment, we propose to employ time reversal in this study to compensate the effect of DS while mitigating the ISI on high frequency data communications. Moreover, the spatial temporal effect provided by TR in both time/space, compresses the undesirable interference on adjacent transceivers by creating a pathway to parallel wireless transmissions while sharing the same time/frequency resources.

%Inspired by aforementioned work, in this paper, we explore time reversal in a high revereberant chip-package focused on both EM and signal transmissions point of view. As complex transceivers would hinder the performance of WNoC architecture, in this work we adopt simple traditional modulation schemes to analyse the performance of the TR signal transmissions in high data rates. 

%\hl{REMAINDER}
The reminder of this paper is organized as follows. Section \ref{sec:background} offers a characterization of the simulation scenario to be employed in this work. It also contains a detailed description of the TR technique applied to flip-chip package as a tool for interference compression. In Section \ref{sec:methodology}, the spatiotemporal effect of TR in the chip is assessed. Section \ref{sec:evaluation_channel} explores the impact of using TR on the performance of modulated data streams. In Section \ref{sec:conclusion}, the paper is concluded.

\section{Methods}
\label{sec:background}

\begin{figure}[!t]
\centering
%%\vspace{-0.2cm}
\includegraphics[width=1\columnwidth]{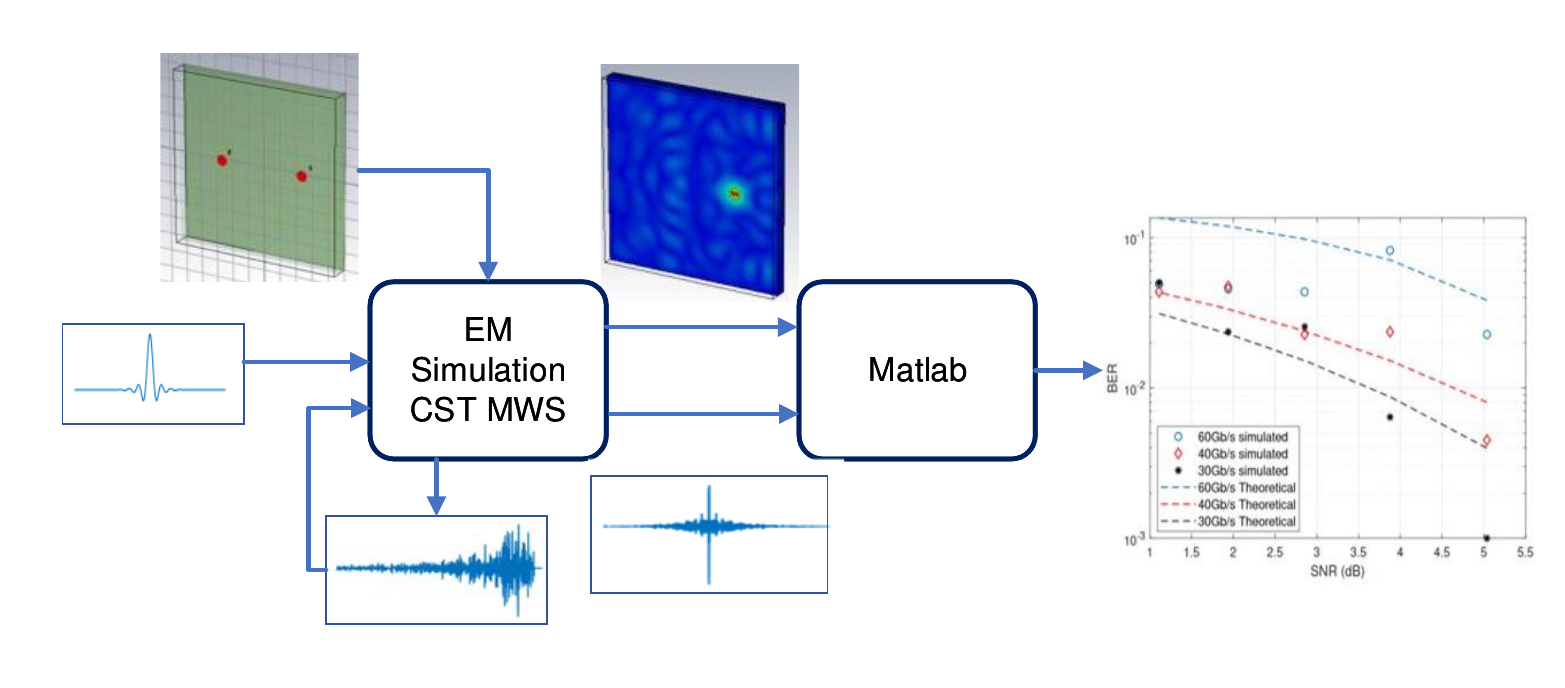}
\vspace{-0.4cm}
\caption{Methodology of this work.}
\label{fig:method}
\vspace{0.4cm}
\end{figure}

Figure \ref{fig:method} summarizes the different methods used in this work. First, we simulate a set of on-chip antennas within a realistic model of a computing package using a full-wave solver, CST Microwave Studio, as described in Sections \ref{sec:environment} and \ref{sec:antenna}. With these simulations, we obtain the impulse response of a particular TX-RX pair. Then, we apply the time-reversal technique as part of the modulation at the transmitter side, as described in Section \ref{sec:TRmethod}.

\subsection{Environment Description}
\label{sec:environment}
%\hl{Explicar package.}
We conduct our simulations in an environment that mimics a flip-chip package, widely used in today's processors. %as shown in \figref{fig:flipchip}.  
In this configuration, the chips are turned over and connected to the system substrate through a set of solder bumps. The packaged chip, therefore, has the silicon substrate on top, which is in turn interfaced by the spreader material and system heat sink on top. The insulator and metal stack are placed at the bottom, interfaced by the solder bumps that connect it to the system \cite{timoneda2018channel, Elmasri2019, chen2019channel}.

The layers are summarized in \tabref{tab:flipParams}. On top the heat spreader, modeled as Aluminum Nitride, dissipates the heat out of the silicon since it has good thermal conductivity; with $\varepsilon_{r}$=8.6 and $\rho$=0.0003. The insulator is silicon dioxide with $\varepsilon_{r}$=3.9 and $\rho$=0.025. 
The silicon die is usually made of bulk silicon and serves as the foundation of the transistors. This layer has $\varepsilon_{r}$=11.9 and would normally have a low resistivity (10 $\Omega\cdot$cm) to favour the operation of digital transistors, yet we consider a high-resistivity silicon layer as in most on-chip antenna works \cite{zhang2007propagation,kim2011interposer} because it dramatically reduces losses at the cost of turning the package in a reverberant environment. In our case, we model this layer as a lossless silicon. 
Finally, we simulate the interconnects and the bumps as one solid layer of copper \cite{timoneda2018channel}. Our chip has a size of 10$\times$10 mm\textsuperscript{2} and is to be studied at a frequency of 60 GHz, although the methodology can be reproduced at higher frequencies as well. The channel simulations will be conducted in CST Microwave Studio \cite{CST}.

\begin{table}[!t] 
\caption{Exploration Parameters.} 
\label{tab:flipParams}  %\small
\vspace{-0.1cm}
\centering
\begin{tabular}{ccccccc} 
\hline
{\bf Parameter} & {\bf Thickness} & {\bf Materials} & {\bf Units} \\
\hline
Lateral Spacer & - & Vacuum & N/A\\
Heat Spreader & 0.5 & Aluminum Nitride & mm \\
Silicon die & 0.5 & High-res. Silicon & mm \\
Chiplet insulator & 0.01 & SiO\textsubscript{2} & mm \\
Bumps & 0.0875  & Copper &  mm\\
%Frequency & 60 & -& GHz \\
\hline
\end{tabular}
%\vspace{-0.5cm}
\end{table}

%\subsection{Spatiotemporal effect of TR in a chip}
\subsection{Antenna and Link Characterization}
\label{sec:antenna}
\begin{figure}[!t]
\centering
%%\vspace{-0.2cm}
\includegraphics[width=1\columnwidth]{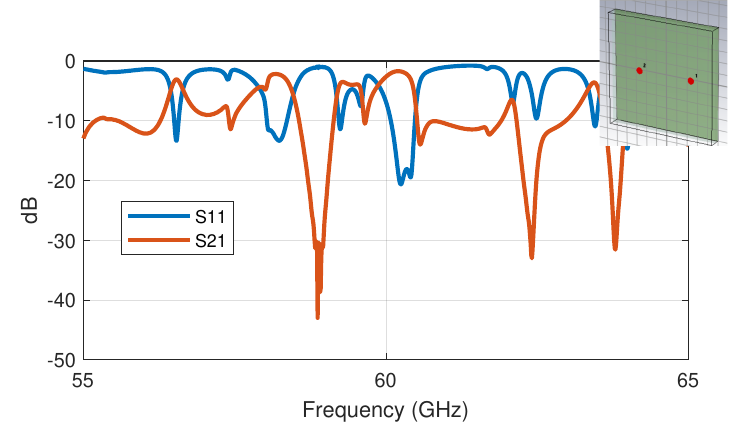}
\vspace{-0.3cm}
\caption{Frequency domain analysis of the proposed antennas and link.}
\label{fig:flipchip_two_monos}
%\vspace{-0.2cm}
\end{figure}

In prior work, the use of a vertical monopole antenna embedded in the silicon layer of the chip has been explored \cite{timoneda2018channel}. We model the antenna as a thin metal cylinder that passes vertically through the silicon layer and is fed by the first metal layers. 
For a 60 GHz frequency, we set the monopole length $L$ to
\begin{equation} \label{equ}
L = \frac{\lambda}{4} = \frac{v_ {p}}{4\cdot f} = \frac{c_0}{4\cdot \sqrt{\varepsilon_{Si}}\cdot f}.
\end{equation} 
where $c_0$ is the speed of light, $f = 60$ GHz is the target frequency, and $\varepsilon_{Si}$ is the permittivity of silicon in that frequency region. Two of the monopole previously described will be placed in the chip at a distance of 7.2 mm from each other. The setup for the simulations, as well as the obtained S parameters, are presented in \figref{fig:flipchip_two_monos}. The S parameters presented are consistent with the presence of high-resistivity silicon, which reduces losses and turns the package into a reverberant environment. This is observed in the resonant nature of the antenna ($S_{11} < -10$ dB at around 60 GHz) and the notchy behavior (but with a moderate average value) of the $S_{21}$. In comparison, the channel spectra in our prior works with lossy silicon \cite{rodriguez2022towards} shows a much worse average value without significant notches.

\subsection{Time Reversal Formulation}
\label{sec:TRmethod}
Time reversal is a pulse compression technique which focuses the received signal energy to its source with back propagation, upon creating a signal reversed in time with prior CSI \cite{lerosey2005time}. The technique is suitable for environments with a long impulse response and, hence, a high delay spread. Though we obtain highly correlated channels in close reverberated spaces such as chip environment, by using the exact CIR data correspond to the source, TR is capable of compressing the co-channel interference in a higher extent with transmissions in same time and frequency while taking the multi-path components on the reflections and reverberation to the advantage of energy focusing.

By assuming each node equipped with a wireless communication interface, let us consider a simple wireless signal transmission from node $A$ to node $B$. Thanks to prior channel characterization, we can assume perfect knowledge of the CIR for such $AB$ link, $h_{AB}(t)$. The time-reversed precoded symbol $x_A$ at node A can be obtained as 
\begin{equation}
x_{A}(t) = h_{AB}(-t) * m_{A}(t),
\label{eqn:trnodeA}
\end{equation}
where $m_{A}(t)$ are the modulated symbols to be transmitted from node $A$ to node $B$, $*$ is the convolution operation, and $h_{AB}(-t)$ is the time-reversed version of the link's CIR, which is
\begin{equation}
h_{AB}(t) = \sum_{k}\alpha_{k} e^{j\psi_{k}}\delta(t-\tau_{k}),
\label{eqn:cnannel}
\end{equation}
with $\{\alpha_k, \psi_{k}, \tau_k \}$ being the amplitude, phase and propagation delay of random multi-path component $k$ of the channel $A\rightarrow B$. $\delta(.)$ denotes the Dirac delta function.

The signal coming from node $A$ that would reach a receiver placed in an arbitrary position $\vec{r}$ can be expressed, in the time domain, as
\begin{equation}
\begin{split}
y(t,\vec{r}) &= h_A(t,\vec{r}) * x_A(t) + n(t) \\
            &= h_A(t,\vec{r}) * h_{AB}(-t) * m_{A}(t) + n(t).
\label{eqn:Rx_enduser2}
\end{split}
\end{equation}
where $h_{A}(t,\vec{r})$ is the CIR of the channel between the position of $A$ and the arbitrary receiver, while $n(t)$ represents the additive white Gaussian noise (AWGN) at the receiver, assumed of zero mean and variance of $\sigma^2$. For instance, the received signal at node $B$ would be
\begin{equation}
y_{B}(t) = h_{AB}(t) * h_{AB}(-t) * m_{A}(t) + n_{B}(t). 
%= \sum_{k}\alpha^2_k \delta(t-2\tau_k)\
\label{eqn:Rx_enduser}
\end{equation}

As can be observed, \eqref{eqn:Rx_enduser} is similar to the usual matched filter expression but only happens when $h_A(t,\vec{r}) = h_{AB}(t)$. In other words, the TR transmission creates a spatial matched filter, this is, a matched filter response only when the receiver is placed (i) in $B$ or (ii) in any other position with the same CIR, which could happen due to spatial symmetries. 

%as exact CIR is known prior to the TR-Processing. In here, a  perfectly matched signal reception could be obtained with the convolution of the TR processed signal with the communication channel ( with similar amplitudes, propagation delays and phase conjugate), thus focusing the energy at receiver end.

%Thus, the received signal at node B with time-reversal can be written as

%As one of the significant advantages of using TR in multi core transmission environment, it was observed as in Figure \ref{TRint} that the interference received on nearby nodes, is  considered to be very low in compared with the signal energy received at intended node with the TR precoded symbol. In here, the interference was analysed by the convolution of the TR precoded symbol with  a set of adjacent channel data obtained from CST within the confined chip environment with 22dB AWGN. The effect of less interference on nearby nodes leads the path way to parallel transmissions on chip environment, by sharing same frequency/time resources. Though the channels could be highly correlated within the reverberant chamber, the rich scattered EM distribution of the received signal compresses the interference in to an acceptable amount, unless the nodes are placed significantly closer to each other.

\begin{figure*}[!ht]
\begin{subfigure}[t]{0.47\columnwidth} 
\includegraphics[width=\textwidth]
{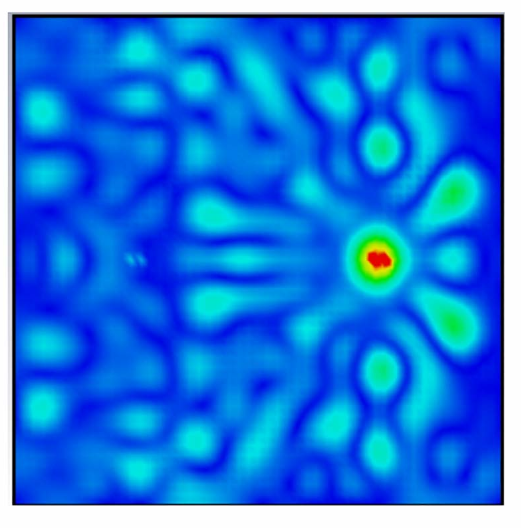}%\vspace{-0.2cm}
 \caption{Field distribution (non-TR).}
\end{subfigure}
\begin{subfigure}[t]{0.65\columnwidth}
 \includegraphics[width=\textwidth]{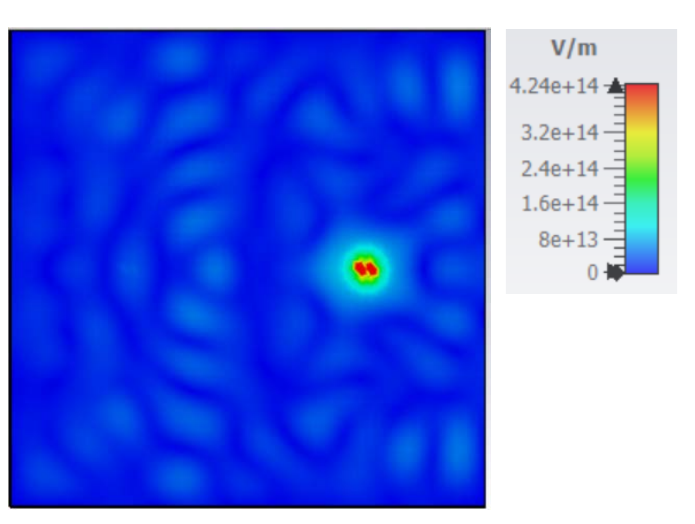}%\vspace{-0.2cm}
 \caption{Field distribution (TR).}
\end{subfigure}
\begin{subfigure}[t]{0.85\columnwidth}
%\vspace{0cm}
  \includegraphics[width=\textwidth]{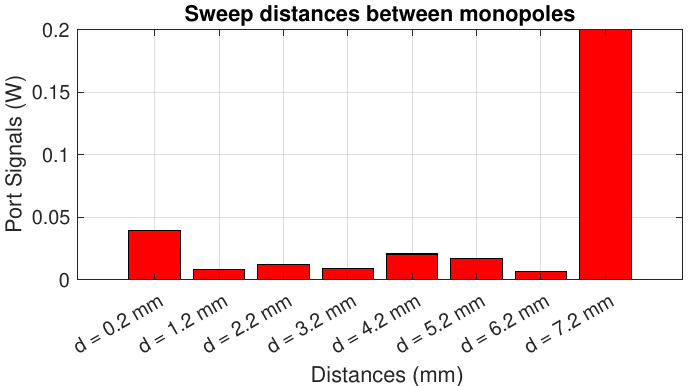}
 \caption{Total received power as a function of distance (TR).}
 \label{fig:distance_sweep}
\end{subfigure}
\vspace{0.3cm}
\caption{Spatial distribution of energy for non-TR and TR transmissions.}
  \label{fig:W_compare_field}
\end{figure*} 

\subsection{Time-Reversal Modulations}\label{TRmod}
As complex transceiver designs are difficult to be incorporated to the WNoC architecture due to high area and power consumption, the basic traditional modulation schemes with simple filters were analysed. In particular, we evaluate Amplitude and Phase Shift Keying (ASK, PSK) and impulse-radio like versions of Pulse Position Modulation (PPM) and On-Off Keying (OOK). The entire analysis is done in MATLAB.
 
For ASK, the signals were multiplied with a continuous carrier wave by using the amplitude ratio of $A1:A2=0.5$ for the bits \{0,1\}, respectively. The BPSK modulation was processed by multiplying the Non-Return to Zero (NRZ) encoded input with a carrier wave. PPM symbols were formed by shifting the impulse radio (IR) on OOK signals with respective prior estimated time shifts.

The data coming from CST was imported in MATLAB and used to create the TR filter as the complex conjugate of $h_{AB}(t)$ and to model $h(t,\vec{r})$ in positions of intended and non-intended receivers. The CST data was then convoluted with the modulated data from the synchronised input signals as in \eqref{eqn:trnodeA}. Thereafter, the processed TR-modulated symbol was transmitted through the channel(s) as detailed in \eqref{eqn:Rx_enduser2}. 

The receiver processing was conducted based on the particular modulation employed in the transmitter end such as (i) synchronised carrier multiplication for BPSK prior to the energy detection and, (ii) instantaneous energy detection with a posteriori threshold calculations based on the received signal energy for ASK and PPM. In case of analysing higher order modulation schemes such as $M$-ary PSK, a similar approach as in BPSK processing can be adopted by multiplying each $I$ and $Q$ phase signal components with the respective synchronised carriers.

%%%%%%%%%%%%%%%%%%%%%%%%%%%%%%%%%%%%%%%%%%%%%%%%%%%%%%%%%%%%%%%%%%%%%%%%%%%%%%%%
%\section{Spatiotemporal effect of TR in a chip}
\section{Impact of Time Reversal on Channel Response}
\label{sec:methodology}
In this section, we show the results obtained in the full-wave solver before and after applying time reversal. As described above, the TR process starts by exciting the transmitting antenna at node $A$ with an impulse in the full-wave solver. The channel response is recorded in the time domain at the position of node $B$, and then time reversed using post-processing. 
%Since this signal somewhat finishes abruptly at 4ns, when applied to the corresponding antenna it does not  show the whole response of the channel of interest. Therefore after time reversing the response of antenna 2, we padded it with zeros up to 20ns to get a clear view of what happens in the channel. 
The time-reversed signal is then fed to node $A$, which radiates it throughout the chip package, but targets node $B$. Next, we analyze the impact of TR in the spatial response, Section \ref{sec:spatial}, and in the temporal response, Section \ref{sec:temporal}.

\subsection{Spatial Focusing}
\label{sec:spatial}
In order to grasp the impact of TR on the spatial distribution of energy, \figref{fig:W_compare_field} shows the field distribution for each transmission at the time of higher concentration of energy at the intended receiver. On the one hand, for the non-TR transmission, the energy goes all over the chip, illustrating the reverberant quality of the chip package. Such an effect would cause high ISI and also hinder the creation of concurrent spatial channels. On the other hand, the spatial focusing of the TR transmission is clear, showing a much higher energy level at the targeted location than anywhere else in the chip. Residual energy can be observed in other locations; this is possible if the $h(t)$ between the transmitter location and an arbitrary position is similar to that of the $A \to B$ link. Yet in our example, the spatial energy scattering is very low compared to the level of spatial focusing achieved by the peak. Hence, we show that the TR technique manages to compensate our highly reverberant channel.    

% \begin{figure}[h]
%   \centering
% \includegraphics[width=\columnwidth]{images/distance_sweep.pdf}
%     \vspace{-0.5cm}
%     %\caption{Time compression TR}
%     %\vspace{-5cm}
%   %\vspace{0.5cm}
%   \caption{Normalized field concentration}
%   \label{fig:distance_sweep}
% \end{figure}

Additionally, we offer a comparison of the difference in concentration on energy when transmitting with a time reverse $h(t)$ towards a non-desired location. To put this to the test, in the landscape shown in \figref{fig:flipchip_two_monos}, we introduce two monopoles separated by 7.2 mm. This is meant to be our default distance and we calculate and time reversed the $h(t)$ for the link formed when the monopoles are in this default position. Then, keeping the transmitting antenna fixed, we change the position of the receiving antenna, applying a displacement in distance bringing them closer. This will form a different link and therefore a different channel. For each of those, we excite the transmitter with the time reversed $h(t)$ obtained for the default distance.

The results regarding the energy concentration are illustrated in \figref{fig:distance_sweep}, which shows the total power received in the different positions. In comparison with the default distance of 7.2 mm, the power received in other distances is very low, meaning that the position of the antenna in the chip will influence the channel response. This proves that we can have a transmission in the intended destination with a high concentration of energy that can be ignored in other spots that are of no interest for the transmission. This is an important conclusion because to have the spatiotemporal focusing that comes with TR we need to use the specific $h(t)$ inherent to each link, giving us the opportunity to have simultaneous transmissions of point-to-point for different links with good concentration and little interference.

\subsection{Temporal Focusing}
\label{sec:temporal}
\begin{figure}[!ht]
    \begin{subfigure}[t!]{\columnwidth}
    \includegraphics[width=\textwidth]{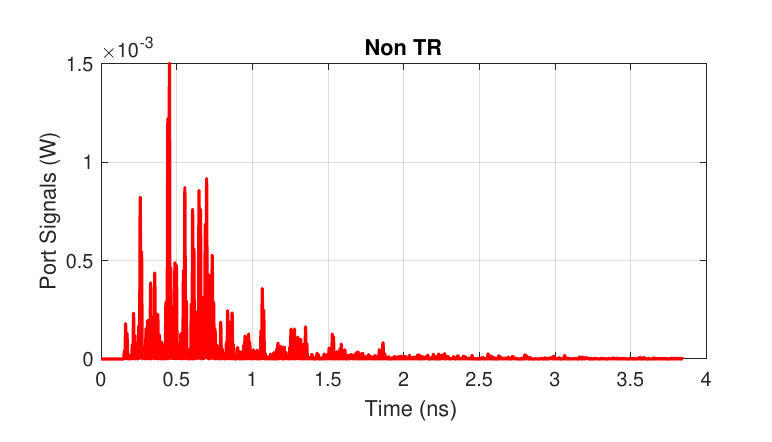}
    %\vspace{-0.5cm}
    \caption{Received signal (non-TR).}
    \label{fig:o21_no_tr}
    %\vspace{-5cm}
  \end{subfigure}
  \vspace{0.5cm}
  \hfill
  \begin{subfigure}[t!]{\columnwidth}
    \includegraphics[width=\textwidth]{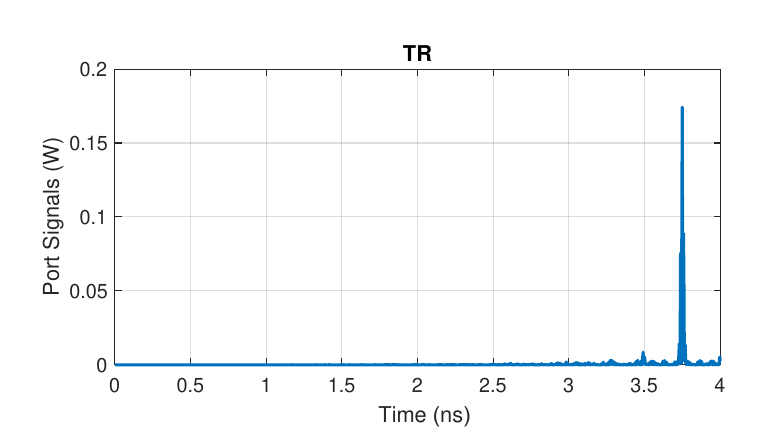}
    %\vspace{-0.5cm}
    \caption{Received signal (TR).}
    %\vspace{-5cm}
    \label{fig:o21_tr}
  \end{subfigure}
  %\vspace{0.2cm}
  %\hfill
  \caption{Temporal energy distribution for non-TR and TR transmissions.}
  \label{fig:W_compare}
\end{figure}

\figref{fig:W_compare} shows a comparison of the signal received in antenna 2 without using TR and when we applied it. For a fair comparison, the \emph{time-reversal filter} is normalized so that the radiated power at antenna 1 is the same in both cases.
%The time signals obtained for these transmissions have been normalized with regard to the power that leaves antenna 2 when applying the signals to it. 
From this figure, we can first observe that the amplitude of the focusing peak is much higher, around three orders of magnitude, than the that of the non-TR signal. It is also clear that the energy is concentrated and not spread over time. 

%\hl{distance comparison show comparison between TR and No TR and distance sweep. Remove figure 7 multicore transmission take it out. Journal paper later}

\begin{figure}[!h]
  \centering
\includegraphics[width=\columnwidth]{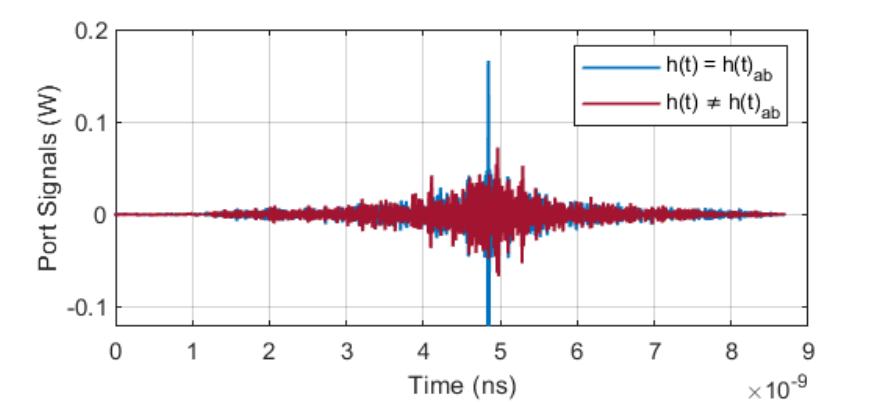}
  \caption{Spatiotemporal focusing: received signal at the intended and non-intended antennas.}
  \label{TRint}
\end{figure}

Finally, to illustrate jointly the spatiotemporal effect of TR, we evaluate the received signal over time both in our targeted antenna 2 and in another antenna placed in an arbitrary close location. As Figure \ref{TRint} shows, the peak is clearly visible in the target antenna, whereas a level around an order of magnitude lower is received in the other nearby antennas. 

%that the interference received on nearby nodes, is  considered to be very low in compared with the signal energy received at intended node with the TR precoded symbol. In here, the interference was analysed by the convolution of the TR precoded symbol with  a set of adjacent channel data obtained from CST within the confined chip environment with 22dB AWGN. 
The effect of less interference on nearby nodes paves the way toward parallel transmissions on the chip environment, by sharing same frequency/time resources and a different \emph{time-reversal channel}. Though the channels could be highly correlated within the reverberant chamber, the rich scattered EM distribution of the received signal compresses the interference to an acceptable amount, unless the nodes are placed significantly closer to each other.

%From this, it is evident that the concentration of energy for the TR transmission is stronger and less scattered. 

%As one of the significant advantages of using TR in multi core transmission environment, it was observed as in Figure \ref{TRint} that the interference received on nearby nodes, is  considered to be very low in compared with the signal energy received at intended node with the TR precoded symbol. In here, the interference was analysed by the convolution of the TR precoded symbol with  a set of adjacent channel data obtained from CST within the confined chip environment with 22dB AWGN. The effect of less interference on nearby nodes leads the path way to parallel transmissions on chip environment, by sharing same frequency/time resources. Though the channels could be highly correlated within the reverberant chamber, the rich scattered EM distribution of the received signal compresses the interference in to an acceptable amount, unless the nodes are placed significantly closer to each other.

\section{Impact of Time Reversal on Communication Performance}
\label{sec:evaluation_channel}
With the superior spatiotemporal focusing observed with a full-wave solver, it is worthwhile to explore the impact on the communications performance. With the extended delay spread due to reverberation as observed in \figref{fig:o21_no_tr}, it is difficult to fit modulated signal transmission as the received signal is disrupted in high-rate transmissions due to ISI. Thus as mentioned in Section~\ref{TRmod}, TR-modulated signal transmissions were analysed to access the performance, importing the $h(t)$ from the full-wave solver. 

\subsection{Rate Analysis}
\begin{figure*}[h]
\begin{subfigure}[t!]{0.33\textwidth}
 \centering
 \includegraphics[scale=0.42]{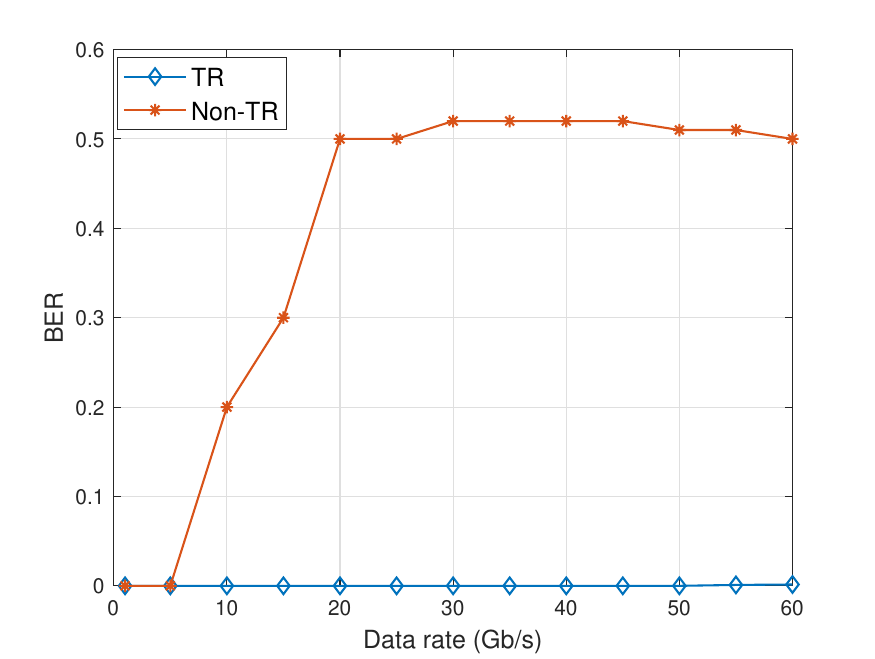}
 \caption{BER vs Data Rate: CW-ASK}
\end{subfigure}%
\begin{subfigure}[t!]{0.33\textwidth}
 \centering
 \includegraphics[scale=0.42]{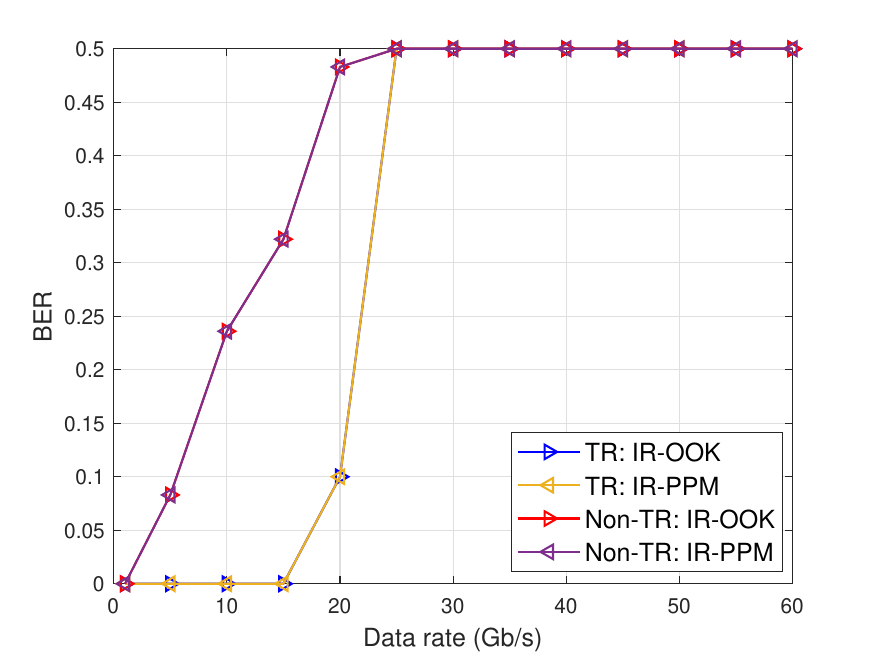}
 \caption{BER vs Data Rate: IR-PPM and IR-OOK}
\end{subfigure}%
\begin{subfigure}[t!]{0.33\textwidth}
 \centering
 \includegraphics[scale=0.42]{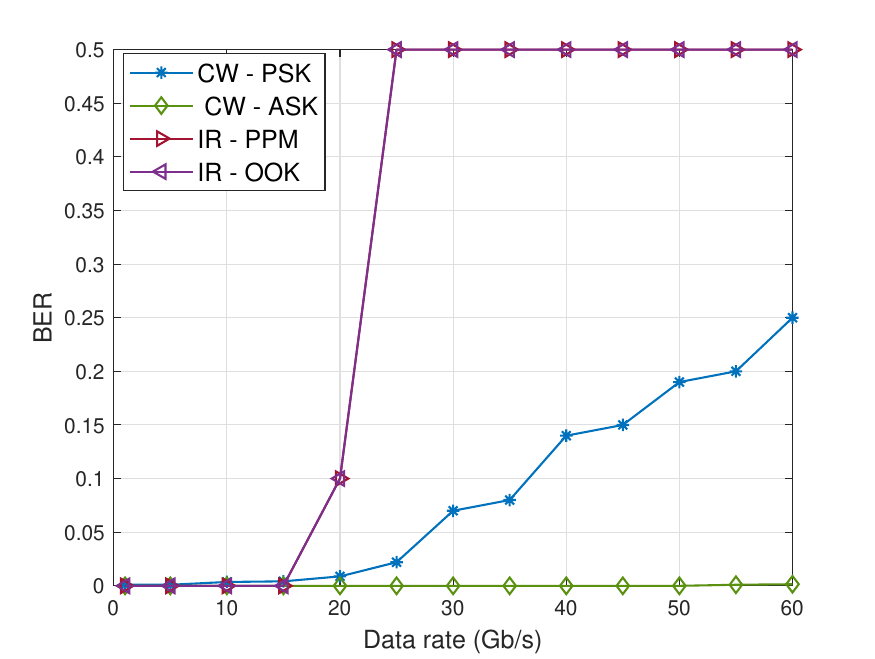}
 \caption{BER vs Data Rate: Modulations with TR}
\end{subfigure}%
\vspace{0.3cm}
  \label{modA}
\begin{subfigure}[t!]{0.33\textwidth}
 \centering
 \includegraphics[scale=0.43]{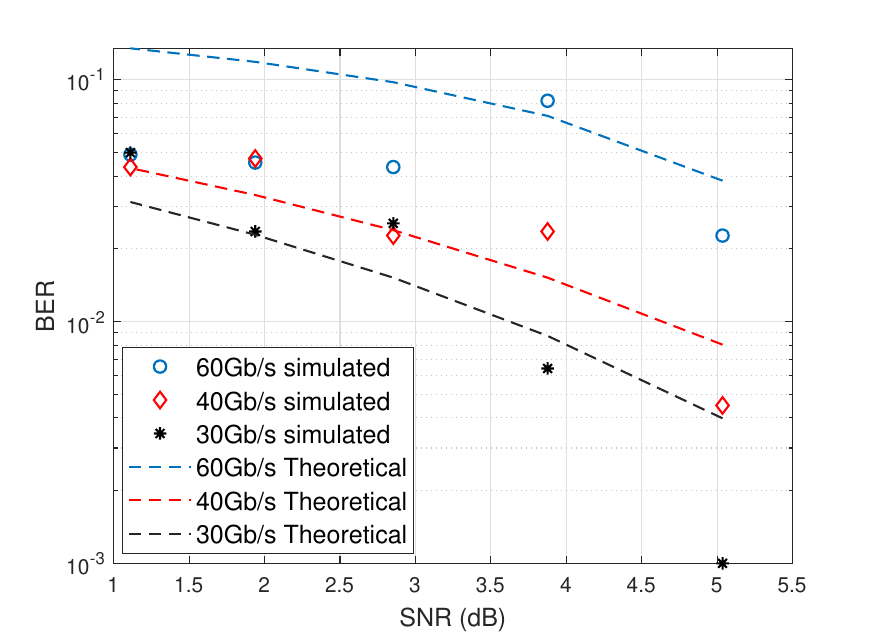}
 \caption{BER vs SNR: CW-ASK}
\end{subfigure}%
\begin{subfigure}[t!]{0.33\textwidth}
 \centering
 \includegraphics[scale=0.36]{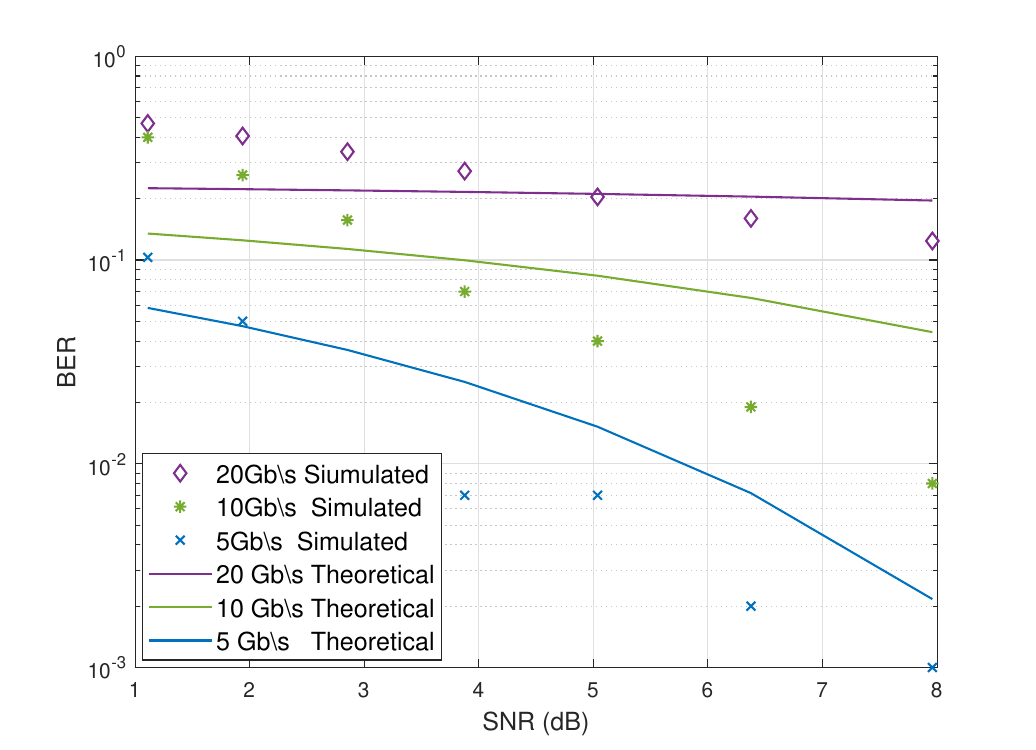}
 \caption{BER vs SNR: IR-PPM}
\end{subfigure}%
\begin{subfigure}[t!]{0.33\textwidth}
 \centering
 \includegraphics[scale=0.42]{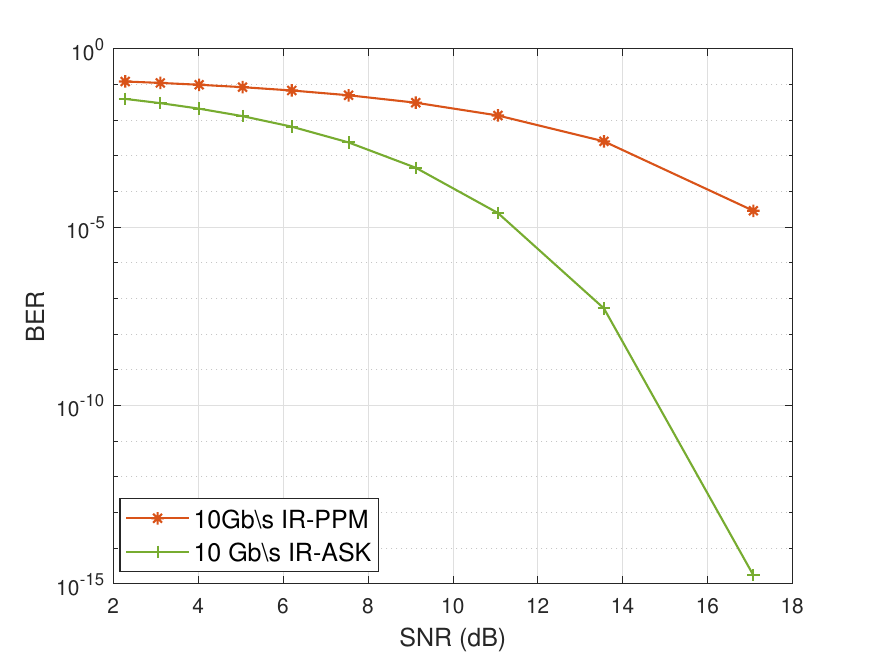}
 \caption{BER vs SNR: Exploring BER with high SNR}
\end{subfigure}%
\vspace{0.3cm}
\caption{Performance of TR with modulations. Unless noted, the SNR is 16 dB and data rate is 10 Gb/s.}
  \label{mod1}
\end{figure*}

The BER of obtained for PSK, ASK, PPM for both impulse radio (IR) and continuous wave (CW) transmissions with 16dB constant SNR and increasing rates can be observed in Figure \ref{mod1}(a)-(c). 

It is shown that, at constant SNR, ASK attains a superior performance with TR by reaching 60 Gb/s with approximately zero BER (limited by the number of simulated bits), while for the non-TR transmissions the performance sharply decreases beyond 5Gb/s. For PSK, a mediocre performance is observed in compared with ASK, but compared with non-TR transmissions, the ISI is at a very low rate for until the data rate reaches 20 Gb/s. PPM and IR-OOK is with equal acceptable performance rates with TR though in PPM multi-node data multiplexing modulation scheme could be introduced in time/frequency with carefully designed additional time shifts for each node. Although not shown for the sake of brevity, it was observed that approximately around 1-5 Gb/s data rates, higher-order modulations such as QPSK/8-PSK has also shown promising results with TR processing. 

\subsection{Power Analysis}
The BER performance of ASK and PPM is further analysed with SNR by maintaining fixing the noise power and increasing the transmitted power. As the simulation measurements were limited with the time consumption on single round cycle detection on higher SNR values, a theoretical BER expression was derived for ASK and PPM bit detection with respect to amplitude shift of scattered received bits. The centroid of the each received bit cluster with 1000 bits was measured and the shift of amplitude ($\Delta A$) was computed. By assuming equiprobable symbols $\{0,1\}$, then 
\begin{equation}
BER_{theoretical}= \frac{1}{2}Q(\frac{\Delta A}{\sigma})\,
\end{equation}
with $\sigma$ being the standard deviation of AWGN and $Q(\cdot)$ being the Q-function. The formulation above was used with the measured amplitude shifts to explore the BER values below 10\textsuperscript{-3} to examine whether the performance with modulations would be compatible with the stringent BER requirements of WNoC. 

As shown in Figure \ref{mod1}, even with the presence of noise, ASK performed very well with 10 Gb/s bandwidth having $10^{-15}$ BER at 17dB. While not shown in the figure, at 60 Gb/s the BER value increases to around 0.02 for ASK. 
Though the BER performance of PPM was not as well as ASK, an acceptable decrement of values, approximately $10^{-5}$ in 17dB can be observed in Figure \ref{mod1}, which could positively lead to lower BER values with increment of signal power. As increasing data rates place the bits closer upon reception, it is hard to adopt PPM beyond 15Gb/s. Furthermore, it was observed with simulations an early BER saturation with a lower performance for PSK comparatively with other modulations. Therefore, the theoretical analysis was limited to ASK and PPM.

\begin{comment}
As TR-ASK provided better performance in compared with PPM and PSK, the BER of ASK was further measured by varying the transmission channel, while keeping a constant TR precoded signal. The measurements were obtained with three different channels separated with three non-equal distances from the source, where $\rho=\frac{\text{The distance from the source to channel data used for TR precoding}}{\text{The distance from source to receiver}}$.  The intend was to observe the spatial temporal focusing under various channel correlations and how the signal superposition could affect the demodulation of the neighbouring cores in high frequency data transmissions. It was observed that ASK-BER showed very low values in nearest cores and comparatively high values in cores placed far from source in compared with the TR-precoded channel distance. Therefore, if concurrent transmissions are to be investigated, careful multi-core paring should be done prior to the data transmission, to mitigate the effect of interference.

\begin{figure}[H]
 \centering
 \includegraphics[scale=0.4]{images/berdist.png}
\caption{Performance of TR with ASK in a variety of channel conditions: BER vs Data Rate}
  \label{mod2}
\end{figure}

\end{comment}
\section{Related Work}
\label{sec:related}
\noindent
%\textbf{Time Reversal in Other Environments}
%\hl{give more details of the paper discussed in introduction an reiterate why are we different from this}
\textbf{Time Reversal in Other Environments.} Time reversal is a technique that has been researched in a variety of domains, especially in acoustics \cite{lerosey2004time}. %This technique consists in sending a short pulse through the medium and capturing it by a time-reversal mirror (TRM).
%The recorded signals are then time-reversed and transmitted back by the TRM. The wave is found to converge back  to its source with a spatiotemporal focusing. 
%It has been found that a complex medium helps the time reverse wave to converge back to the source more accurately \cite{lerosey2004time}. 
The possibility of transposing the idea of TR to wireless communications is also discussed and proven at microwave frequencies in \cite{lerosey2005time}. They argue that a TR antenna will compensate for the multipath that comes with the large-scale wireless channel and also be able to increase the transmission rate as a result of the reverberation of the medium. There has been also research on applied TR at very high frequencies. In
\cite{mokh2022experimental}, the authors propose to exploit the diversity of the channel and assess the performance of TR technique at a carrier frequency of 273 GHz. Their multipath channel for this approach was created using an aluminum tube as waveguide. An extension of this work is presented in \cite{highfreqTR}, which shows spatiotemporal focusing in three different scenarios also including subTHz bands. Besides the previous examples, some experimental studies \cite{TRPPM1,TRPPM2} have also explored the combination of the TR technique with multiple modulations. The main difference of our work with respect to the previous examples resides in the environment, which is a chip package instead of larger setups emulating indoor environments, for instance. Another difference is the use of the non-static conventional version of TR, where a two-way approach is required, i.e. frequent probing to obtain the CIR and then the actual application of TR. 
%Our approach consists in applying TR transmission at the chip scale. This is a naturally reverberant environment in which 
Instead, we can take advantage of the %fact that the landscape is highly controlled and known beforehand by the designer. This makes for a 
quasi-deterministic and time-invariant channel to pre-calculate the CIR only once. 
%. Therefore the CIR of the engineered medium can be obtained with prior channel characterization and will remain the same for that particular link.

%Because the TR technique thrives in a dispersive channel, most of the work developed in TR related to EM includes some sort of reverberation chamber, a metallic box around the setup, or multiple metallic rods to enable the scattering of waves.
%Our approach consists in applying TR transmission at the chip scale. This is a naturally reverberant environment in which  we can take advantage of the fact that the landscape is highly controlled and known beforehand by the designer. This makes for a quasi-deterministic and time-invariant channel. Therefore the CIR of the engineered medium can be obtained with prior channel characterization and will remain the same for that particular link.

\textbf{Arrays and Spatial Multiplexing.} An alternative to TR to increase the capacity of the wireless network is to use arrays to obtain spatial channels through highly directive transmissions. 
%This includes MAC protocols to share the channel among the cores and  multiplexing schemes in frequency, time, and code. 
In the context of chip-scale communications, beam forming and beam switching have been studied in multiple works. In \cite{Baniya2018, Narde2020}, authors use planar arrays at 60 GHz. In \cite{Obeamforming}, authors take advantage of the proximity of antennas in multiple processors of the chip and exploit the existing infrastructure to form antenna arrays in a dynamic and opportunistic fashion. More recently, in \cite{rodriguez2022towards}, a compact phased array is proposed to achieve concurrent multicast channels at 60 GHz and 110 GHz. All these works, however, consider either unpackaged configurations (i.e. open-die) \cite{Baniya2018,Narde2020} or flip-chip with lossy silicon \cite{Obeamforming,rodriguez2022towards}, in which cases the DS is limited. However, processors are typically fully packaged as we consider in this work, which either limits the efficiency due to losses or the speed due to DS. Here, we achieve both speed and efficiency thanks to TR.

%The presented work in \cite{Obeamforming, rodriguez2022towards} take place in a chip package with the same features as the one used in this paper, this means that the silicon layer is modeled as bulk silicon for the operation of the transistors. The high permittivity of the silicon layer allows forming the compact antenna arrays proposed in \cite{rodriguez2022towards} with little coupling effect among antennas. These scenarios will perform poorly for TR, because the high losses provided by the bulk silicon will attenuate part of the signal so there will be fewer propagation phenomena going on in the chip and there's a direct relationship between the multipath richness of the environment and the efficiency of TR.

%\noindent
%\textbf{Wireless Links within Computing Packages}

%\noindent
%\textbf{Arrays and Spatial Multiplexing}
%\hl{cite 2022 my paper and opportunistic arrays and say why we cant have TR there}

\section{Conclusion}
\label{sec:conclusion}

\begin{comment}
In this paper, we have conducted several simulations to explore the TR technique in an enclosed package. 
Our first result is the confirmation that TR can be used on a computer chip to take advantage of its high reverberation and create temporal and spatial effects. 
To test the quality of these effects a performance analysis has been made. This study was done comparing the BER for different modulations with variable data rates, both for transmission with and without TR. Overall, the transmissions using TR with the increase in the data rate exhibited a lower BER. ASK-TR modulation was found to have the best performance by reaching data rates of 60 Gb/s with a perfect BER.
\end{comment}
In this study, we analysed the performance of WNoC with a simplified version of time-reversal signal transmission on a flip-chip package. Initial experiments were conducted to evaluate the spatiotemporal focusing effect with a full-wave solver. Then, we combined time-reversal with simple modulations at high frequencies and analyzed its performance. It was demonstrated that with ASK/PSK/PPM pre-coded with time reversal provides acceptable bit error rates beyond 10 Gb/s. ASK provided the best BER performance with $10^{-15}$ at 10 Gb/s in 17dB AWGN, by complying with the stringent BER requirements of the scenario. Furthermore, the interference compression exhibited with TR upon transmitting signals suggests that parallel concurrent signal transmissions can be achieved within the same time/frequency channel with minimum interference. In future work, we aim to explore such multiplexing capabilities as well as the impact of non-ideal filters on the performance of the TR technique.

\section*{Acknowledgment}
Authors acknowledge support from the European Union’s Horizon 2020 research and innovation program, grant agreement 863337 (WiPLASH), the European Research Council (ERC) under grant agreement 101042080 (WINC) as well as the European Innovation Council (EIC) PATHFINDER scheme, grant agreement No 101099697 (QUADRATURE).

%%% Local Variables:
%%% mode: latex
%%% TeX-master: "main"
%%% End:

\bibliographystyle{ACM-Reference-Format}
\bibliography{bib2}

% % --- Appendix ---%
\appendix
%\section{Overflow form other sections}
%\label{sec:set-diff-dodis}
%Sometime you ware super excited about some details that does not quite fit with
%the rest of the paper goes here. For example, some details about how you
%instrumented the Android Linux kernel should go to appendix, and for really
%curious reader to read. Remember it's appendix, so the reader is not required to
%read, and you should not put critical information in appendix that is crucial
%for understanding the rest of the paper.

%%% Local Variables:
%%% mode: latex
%%% TeX-master: "main"
%%% End:

\end{document}